\documentstyle[11pt]{article}

\newcommand{\msbar}{{\rm \overline{MS\kern-0.14em}\kern0.14em}}

\begin{document}

\begin{flushright}
Liverpool Preprint: LTH 351\\
 hep-lat/9509090\\
26 June 1995\\
\end{flushright}

\vspace{5mm}

\begin{center}
{\LARGE\bf
Hadronic Physics from the Lattice
}\\[10mm]

{\bf  Chris Michael\footnote{
presented at the  NATO Advanced Study Institute on
Hadron Spectroscopy and the Confinement Problem
                     June 27 - July 7, 1995}
  }\\

{DAMTP, University of Liverpool, Liverpool, L69 3BX, U.K.}\\
\end{center}

\begin{abstract}

We present the lattice gauge theory approach to evaluating
non-perturbative hadronic interactions from first principles.   We
discuss applications to glueballs, inter-quark potentials, the running
coupling constant, the light hadron spectrum  and the pseudoscalar decay
constant $f_P$.

\end{abstract}

\section{Introduction}

The theory of the strong interactions is accurately provided by Quantum
Chromodynamics (QCD). The theory is defined in terms  of elementary
components: quarks and gluons. The only free parameters  are the quark
mass values (apart from an overall energy scale imposed  by the need to
regulate the theory). This formulation is essentially the unique
candidate for the theory of the strong interactions. The only feasible
way to  describe a departure from QCD would be in terms of quark and
gluon  substructure.  At least at the energy scales up to which it is
tested, there  is no evidence for such substructure.

QCD provides a big challenge to theoretical physicists. It is defined
in terms of quarks and gluons but the physical particles are
composites: the mesons and baryons. Any complete description must  then
yield these bound states: this requires a non-perturbative  approach.
One can see the limitations of a perturbative approach by  considering
the vacuum: this will be approximated in perturbation theory as
basically empty with  rare quark or gluon loop fluctuations. Such a
description will allow  quarks and gluons to propagate essentially
freely which is not the case experimentally. The true (non-perturbative)
vacuum can be better thought of as a  disordered medium with whirlpools
of colour on different scales. Such a  non-perturbative treatment then
has the possibility to explain  why quarks and gluons do not propagate
(ie quark confinement).

The main contender for a non-perturbative description of QCD is the
lattice gauge theory approach. Following the ideas of Wilson, space-time
is replaced by a discrete grid (the lattice) but gauge invariance is
retained exactly. Using also periodic boundary conditions in space and
time, the system  will have a finite number of degrees of freedom:  the
gluon and quark fields at the lattice sites. Actually gauge invariance
implies that the gluon field lies on the links of the lattice (they
correspond to the  mechanism that allows colour orientation at different
sites to be  compared). This finite number of degrees of freedom implies
that the theory is a quantum many-body problem rather than a field theory.

The step which makes this many-body problem tractable is to consider
Euclidean time. This is perhaps the step which is the most difficult
to assess. Formally the quantities to be evaluated can be expressed
as Green functions: vacuum expectation values of products of fields.
The existence and properties of these Green functions under a Wick
rotation to Euclidean time are widely used in perturbative treatments.
A non-perturbative determination of the Green functions in
the Euclidean time case will lead directly to quantities of relevance
in the physical Minkowski time case (eg masses and matrix elements).

With the Euclidean time approach, the formulation of QCD (consider, for
example, the  functional integral over the gauge fields) is converted
into a  multiple integral which is well defined mathematically. For a
lattice of $L^4$ sites with a colour gauge group of SU(3), this  would
be a $8 \times  4 \times L^4$ dimensional integral (8 from  the colour
group manifold, $4L^4$ from the gauge fields on each link). For any
reasonable value of $L$,  this is a very high dimension indeed.
Simpson's rule is not the way  forward!  The standard approach is to
use a Monte Carlo approximation  to the integrand.  This is implemented
in an ``importance sampling" version  so that a stochastic estimate of
the integral is made from a finite  number of samples (called
configurations) of equal weight. The construction  of efficient
algorithms to achieve this is a topic in itself. Here I  will
concentrate on the analysis of the outcome, assuming that such
configurations have been generated.

So what we have at our disposal is a set of samples of the vacuum. It
is then straightforward to evaluate the average of various products  of
fields over these samples - this gives the Green functions by
definition.  The Green function can then be continued from Euclidean to
Minkowski time (in most cases this  is trivial) and compared to
experiment.

The validation of the lattice approach calls for a series of checks
that everything is under control.
 \begin{itemize}
 \item the lattice spacing should be small enough (discretisation
errors)
 \item the lattice must be big enough in space and time (finite size
errors)
 \item the statistical errors must be under control
 \item Green functions must be extracted  with no contamination (eg a
ground state mass could be contaminated  with a piece coming from an
excited state)
 \item both the quark contribution to the vacuum (sea quarks)  and the
quark constituents of hadrons (valence quarks) are  usually treated by
using larger mass values than the  experimental ones and then
extrapolating. This extrapolation must be treated accurately.
 \item a particularly severe approximation is to treat the sea-quarks as
of infinite  mass. This corresponds to neglecting quark loops in the
vacuum and is computationally very much faster. This is known as the
``quenched approximation''.
 \end{itemize}

The most subtle of these is the discretisation error. In order to
extract the continuum limit of the lattice, one must show that the
physical results will not change if the lattice spacing is decreased
further.  This is subtle because the lattice spacing is not known
directly - in effect,  it is measured. The lattice simulation is
undertaken at a value  of a parameter conventionally called $\beta$. In
the limit of small  coupling $g^2$, where perturbation theory applies,
$\beta=6/g^2$. Thus  large $\beta$ corresponds to small $g^2$. Now,
perturbatively,  the coupling $g^2$ corresponds to the  lattice spacing
$a$  as
 $$
g^2\approx {1.0 \over b_0 \log a^{-2}} \ \ \hbox{where}\ \
b_0={11-{2 \over 3} N_f \over 16\pi^2}
$$
 for $N_f$ flavours of quarks. For the case of interest,
$N_f \le 3$, this corresponds to small values of $g^2$ at small distance
scale $a$ -  as expected from  asymptotic freedom.

The perturbative argument is appropriate  to the study of results at
large  $\beta$ (small $g^2$). We will find that lattice simulation  of
QCD uses values of $\beta \approx 6$ and hence bare couplings
corresponding to $\alpha = g^2/(4\pi) \approx 0.08$. In the pioneering
years of lattice work, this was thought to be a sufficiently small
number that the perturbation series would converge rapidly. One of  the
major advances, in recent years, has been the realisation that  the bare
lattice coupling (our $g^2$ above) is a very poor  expansion parameter
and the perturbation series in the bare coupling does not converge well
at the $\beta$ values of interest.  The theoretical explanation for this
poor convergence is that the  lattice Lagrangian differs from the
continuum Lagrangian and allows extra  interactions. These include
tadpole diagrams~\cite{LM} which have the property that  they sum up to  give a
contribution that involves high order terms in the  perturbation series.
The way to avoid this problem with tadpole terms is to  use a
perturbation series in terms of a renormalised coupling -  rather than
the bare lattice coupling. I return to this topic  when discussing the
lattice determination of  $\alpha_S$ later.

This change of attitude to the method of determining $a$ from  $\beta$
has had considerable implications  for lattice predictions, as I now
explain, since we wish to work in a region of lattice spacing $a$  where
perturbation  theory in the bare coupling is not precise. Because of
this,  in practice, $a$  is determined from the non-perturbative lattice
results themselves.  Thus if the energy of some particle is measured on
the  lattice, it will be available  as the dimensionless combination
$\hat{E}$. From a value for $E$ in physical units,  then $a$ can be
determined since, on dimensional grounds, $\hat{E} =Ea$. Furthermore, by
increasing $\beta$, the  change in the observable $\hat{E}$ gives
information about the change in $a$  since $E$ is fixed - assuming it is
the physical mass.  This should allow  a calibration of $\beta$ in
terms  of $a$ to be established.

This procedure is overly optimistic, however. The discretised lattice
theory is different from the continuum theory on scales of the  order of
the lattice spacing $a$. For the Wilson action formulation  of gauge
theory, this implies that the continuum energy $E_c$ is related  to the
lattice observable $\hat{E}(\beta) = Ea$ as
 $$
 { \hat{E} (\beta) \over a }  = E_c + {\cal O}(a^2)
$$
 A direct consequence is that the ratio of two energies (of
different particles, for example) will have  discretisation
errors of order $a^2$.

Thus, to cope with discretisation errors, the procedure required is
to evaluate dimensionless ratios of quantities of physical interest
at a range of values of the lattice spacing $a$ and then extrapolate
the ratio to the continuum limit ($a \to 0$).

Note that when the fermionic terms are included, the discretisation
error is of order $a$ for the Wilson fermionic action. By adding
further terms in the fermionic action, the error can be reduced -
to order $\alpha a$ for the SW-clover fermion formulation.

\section{Glueball Masses}

I choose to illustrate the workings of the lattice method by describing
the determination of the glueball spectrum.  Of course, glueballs are
only defined unambiguously in the quenched approximation - where
quark loops in the vacuum are ignored. In this approximation, glueballs
are stable and do not mix with quark - antiquark mesons.  This approximation
is very easy to implement in lattice studies: the full gluonic action
is used but no quark terms are included. This corresponds to a full
non-perturbative treatment of the gluonic degrees of freedom in
the vacuum. Such a treatment goes much further than models such as the
bag model.

The glueball mass can be measured on a lattice through evaluating the
correlation $C(t)$ of two closed  colour loops (called Wilson loops) at
separation $t$ lattice  spacings. Formally
 $$
C(t) =
<0|G(0) G^{\dag}(t)|0> = \sum_{i=0} c_i
<g_i(0)|g_i(t)> c^*_i = \sum_{i=0} |c_i|^2 e^{-\hat{m}_i t }
$$
 where $G$ represents the closed colour loop which can be thought  of as
creating a glueball state $g_i$ from the vacuum. Summing  over a
complete set of such glueball states (strictly these are eigenstates  of
the lattice transfer matrix  where $\exp{-\hat{m}_i}$ is the lattice
eigenvalue corresponding to a step of one lattice spacing in time) then
yields the above  expression. As $t \to \infty$, the lightest glueball
mass will dominate.  This can be expressed as
 $$
\hat{m}_0 = \lim_{t \to \infty}\hat{m}_{\rm eff}(t) \ \
\hbox{where} \ \   \hat{m}_{\rm eff}(t) = \log({ C(t-1) \over C(t)})
 $$
Note that since for the excited states $\hat{m}_i > \hat{m}_0$, then
$\hat{m}_{\rm eff}(t) > \hat{m}_{\rm eff}(t+1) > \hat{m}_0$. This
implies that the effective mass, defined above, is an upper bound  on
the ground state mass.  In practice, sophisticated methods are used to
choose loops $G$ such  that the correlation $C(t)$ is dominated by the
ground state glueball  (ie to ensure $|c_0| >> |c_i|$ ). By using a
several different loops,  a variational method can be used to achieve
this effectively.  These techniques are needed to obtain accurate
estimates of $\hat{m}_0$ from  modest values of $t$ since the signal to
noise decreases as $t$ is  increased. Even so, it is worth keeping in
mind that upper  limits on the ground state mass are obtained in
principle.

The method also needs to be tuned to take account of the many
glueballs: with different $J^{PC}$ and different momenta. On the lattice
the Lorentz symmetry is reduced to that of  a hypercube. Non-zero
momentum sates can be created (momentum is discrete in units of
$2 \pi /L$ where $L$ is the lattice spatial size). The usual
relationship between energy and momentum is found for sufficiently
small lattice spacing. Here we shall concentrate on the simplest
case of zero momentum (obtained by summing the correlations over
the whole spatial volume).

For a state at rest, the rotational symmetry becomes a  cubic symmetry.
The lattice states (the $g_i$ above) will transform  under irreducible
representations of this cubic symmetry group (called $O_h$). These
irreducible representations can be linked to  the representations of the
full rotation group SU(2). Thus, for  example, the five spin components
of a $J^{PC}=2^{++}$ state should be appear as  the two-dimensional
E$^{++}$ and the three-dimensional T$_2^{++}$ representations on  the
lattice, with degenerate masses. This degeneracy requirement then
provides a  test for the restoration of rotational invariance -  which
is expected to occur at sufficiently small lattice spacing.

The results of lattice measurements~\cite{DForc,MT,glue,gf11} of the
$0^{++}$ and  $2^{++}$ states are shown in fig~1. The restoration of
rotational  invariance is shown by the degeneracy of the two $O_h$
representations that make up the $2^{++}$ state.  Fig~1 shows the
dimensionless combination of the  lattice glueball mass $\hat{m}_0$ to a
lattice quantity $\hat{r}_0$. We will return to describe  the lattice
determination of $\hat{r}_0$ in more detail - here it suffices to accept
it as a well measured quantity on the lattice that can be used to
calibrate the lattice spacing and so explore the continuum limit. The
quantity plotted, $ \hat{m}_0 \hat{r}_0$, is expected to be equal to the
product of continuum quantities $m_0 r_0$ up to corrections of order
$a^2$. This is indeed  seen to be the case. The extrapolation to the
continuum limit ($a \to 0$) can then be made with confidence.

The value of $r_0$ in
physical units is about 0.5 fm and we will  adopt a scale equivalent to
$r_0^{-1} = 0.372 $ GeV.
 This information yields lattice  predictions
for the glueball masses of around 1.6 GeV and 2.2 GeV  for the $0^{++}$
and $2^{++}$ glueballs respectively.

We return briefly to the independence of the results on the volume  of
the lattice. In the early days of glueball mass determination, it  was
expected that a spatial size $L$ should satisfy  $\hat{m}(0^{++})L > 1 $
and, hence, that values of $\hat{m}(0^{++})L $ of 1 to 4 would suffice.
A careful lattice  study~\cite{smallg}  showed that $\hat{m}(0^{++})L >
8 $ was required to  obtain rotational invariance and a result
independent of $L$. The results  collected in fig~1 all satisfy this
latter inequality so can be regarded as the infinite volume
determination.

\begin{figure}[p]
\vspace{14cm} 
\includegraphics{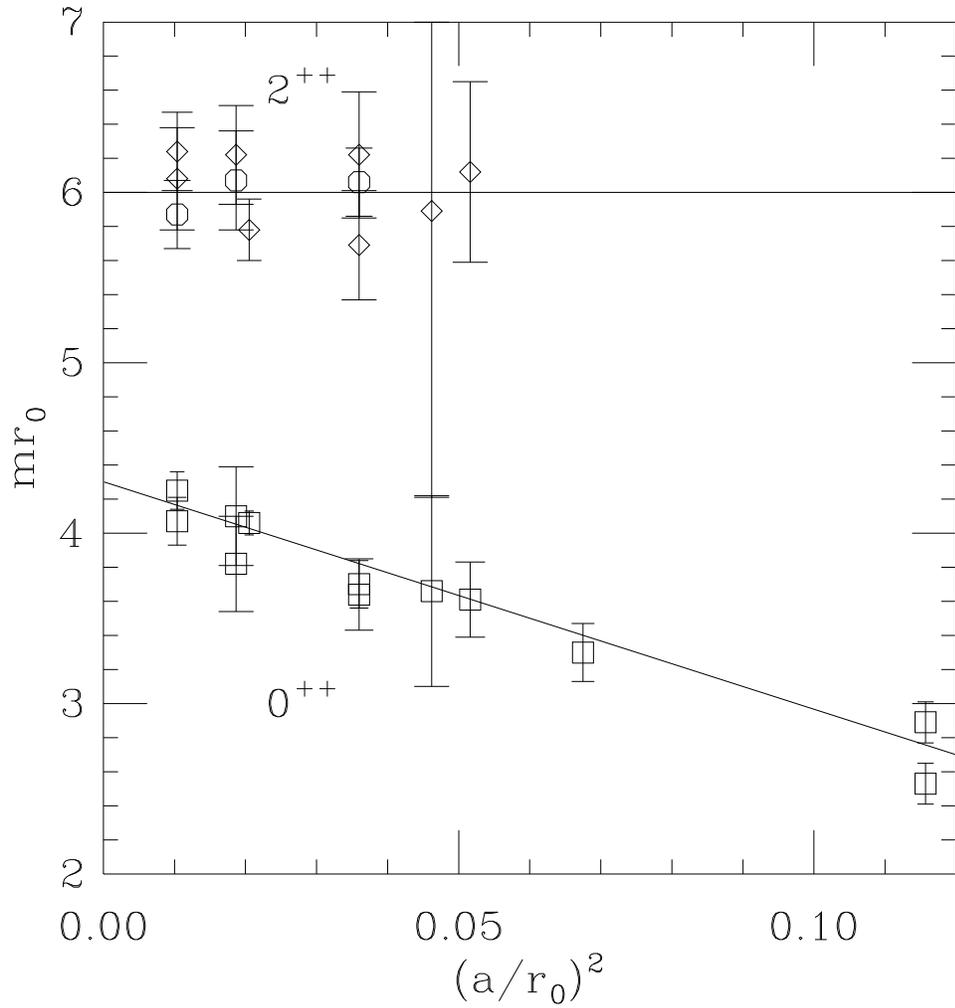}
\caption{ The value of mass of the  $J^{PC}=0^{++}$ and $2^{++}$
glueball states from refs{\protect\cite{DForc,MT,glue,gf11}} in units of
$r_0$. The $T_2$ and $E$ representations are shown  by  octagons  and
diamonds respectively.  The straight line  shows a fit describing the
approach to the continuum limit as $a \to 0$.
   }
\end{figure}

The predictions for the other $J^{PC}$ states are that they lie
higher in mass and the present state of knowledge is summarised
in fig~2.  Remember that the lattice results are strictly
upper limits. For the   $J^{PC}$ values not shown, these
upper limits are too weak to be of use.

Since quark - antiquark mesons can only have certain  $J^{PC}$ values,
it is of special interest to look for glueball with  $J^{PC}$ values
not allowed for such mesons: $0^{--}, 0^{+-}, 1^{-+}, 2^{+-}, $ etc.
Such spin-exotic states, often called ``oddballs'',  would not mix
directly with quark - antiquark mesons. This would  make them a very
clear experimental signal of the underlying glue dynamics.  Various
glueball models (bag models, flux tube models, QCD sum-rule  inspired
models,..) gave different predictions for the presence of such oddballs
(eg. $1^{-+}$) at relatively low masses. The lattice mass spectra
clarify these uncertainties but, unfortunately for experimentalists,  do
not indicate any low-lying oddball candidates. The lightest candidate
is from the T$_2^{+-}$ spin combination. Such a state could correspond
to an $2^{+-}$ oddball. Another interpretation is also possible,
however,  namely that a non-exotic $3^{+-}$ state is responsible (this
choice of interpretation can be resolved in principle by finding the
degenerate 5 or 7 states of a $J=2$ or 3 meson). The overall  conclusion
at present is that there is no evidence for any oddballs  of mass less
than 3 GeV.

Glueballs are defined in the quenched approximation and, hence, they  do
not decay into mesons since that would require quark - antiquark
creation. It is, nevertheless, still possible to estimate the  strength
of the matrix element between a glueball and a pair of mesons  within
the quenched approximation. For the glueball to be a relatively narrow
state, this matrix element must  be  small. A  very preliminary attempt
has been made to estimate the size of  the coupling of the $0^{++}$
glueball to two pseudoscalar mesons~\cite{gdecay}. A relatively small
value is found. Further work needs to be done to  investigate this in
more detail, in particular to study the mixing between  the glueball and
$0^{++}$ mesons since this mixing may be an important  factor in the
decay process.

Another lattice study  will become feasible soon. This is to study
the glueball spectrum in full QCD vacua with sea quarks of mass $m_D$.
For large $m_D$, the result is just the quenched result described above.
For $m_D$ equal to the experimental light quark masses, the results
should just reproduce the experimental meson spectrum - with the
resultant uncertainty between glueball interpretations and other
interpretations. The lattice enables these uncertainties to be resolved
in principle: one obtains the spectrum for a range of values of $m_D$
between these limiting cases, so mapping glueball states at large
$m_D$ to the experimental spectrum at light $m_D$.

\begin{figure}[p]
\vspace{16cm}
\includegraphics{gbspect.fig}
  \caption{ The  mass of the  glueball states  with quantum  numbers
$J^{PC}$ from ref{\protect\cite{glue}}.  The scale is set by
${\protect\sqrt{\sigma}} \approx 0.44 $ GeV which yields the right hand
scale in GeV. The solid points represent mass determinations whereas the
open points  are upper limits.
   }
\end{figure}

\section{Potentials between quarks}

A very straightforward quantity to determine from lattice
simulation is the interquark potential in the limit of very
heavy quarks (static limit). This potential is of direct
physical interest because solving the Schr\"odinger equation in
such a potential provides  a good approximation to the $\Upsilon$
spectrum.  It is also relevant to exploring both confinement
and asymptotic freedom on a lattice.

The basic route to the static potential is to evaluate the average
$W(\hat{R},\hat{T})$ in the vacuum samples of a rectangular closed loop
of colour flux  (a Wilson loop of size $\hat{R} \times \hat{T}$).  here
the lattice quantities $\hat{R}$ and $\hat{T}$ are related to  the
physical distances $R$ and $T$ by $\hat{R}=R/a$, etc where  $a$ is the
lattice spacing which is not known explicitly. Then it can be shown that
the required static potential in lattice units is given by
 $$
\hat{V}(\hat{R}) = \lim_{\hat{T} \to \infty} \log{W(\hat{R},\hat{T}-1)
\over W(\hat{R},\hat{T})}
 $$
The limit of large $T$ is need to
separate the required potential from  excited potentials. This limit can
be made tractable in practice by  using more complicated loops than the
simple rectangular loop described  above.

A summary of results~\cite{pots} for the potential at large $R$ is shown
in fig~3.  The result that the force $dV/dR$ tends to a  constant at
large $R$ (and thus $V(R)$ continues to rise as $R$ increases) is  a
manifestation of the confinement of heavy quarks (in the quenched
approximation). The force  appears to approach a constant at large $R$.
A simple parametrisation  is traditional in this field:
 $$
V(R) = V_0 - { e \over R} + K R
 $$
where $K$ (sometimes written $\sigma$) is the string tension. The
term $e/R$ is referred to as the Coulombic part in analogy to the
electromagnetic case. The equivalent relationship in terms of
quantities defined on a lattice is
 $$
\hat{V}(\hat{R}) = \hat{V}_0 - { e \over \hat{R}} + \hat{K} \hat{R}
 $$
where the lattice string tension $\hat{K}=Ka^2$.

\begin{figure}[th]
\vspace{8cm}
\includegraphics{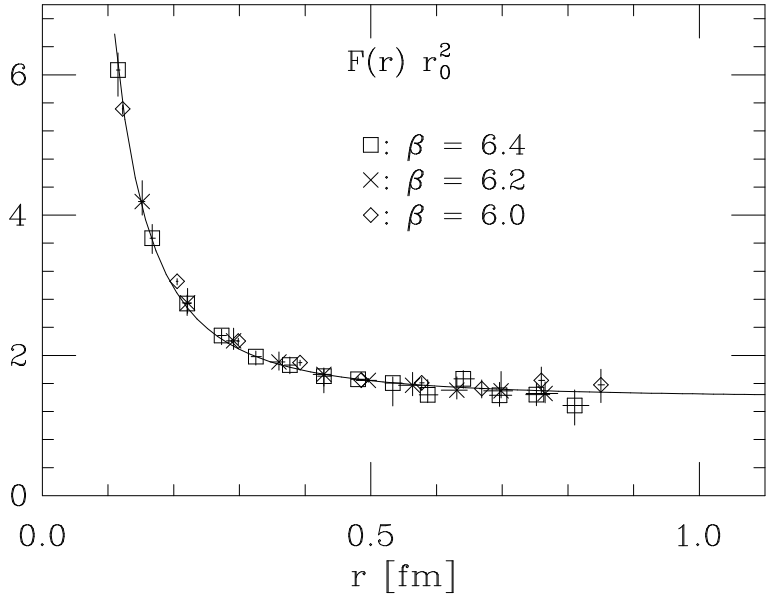}
  \caption{ The force between static quarks at separation $r$
as measured on lattice with different lattice spacing $a$ (with
the $\beta$-values shown).
   }
\end{figure}

Since the string tension is given by the slope of $V(R)$ against  $R$ as
$R \to \infty$, this implies that some error will arise in determining
$K$  coming from the  extrapolation of lattice data at finite $R$. A
practical resolution  is to define a value of $R$ where the potential
takes a certain form. The convention is to use $r_0$ where
 $$
 \left. {\hat{R}^2 {d\hat{V}(\hat{R}) \over d\hat{R}}}
\right|_{\hat{r}_0} = 1.65
 $$
 Thus $\hat{r}_0$ can be determined by interpolation in $\hat{R}$ rather
than  extrapolation.  In practice, this means that $\hat{r}_0$ is very
accurately  determined by lattice measurements and so is a useful
quantity to use  to set the scale since $\hat{r}_0=r_0/a$. With the
simple parametrisation above, we have $r_0^2=(1.65-e)/K$ so $r_0$ is
closely related to the string tension  since $e \approx 0.25$.  The
string tension is usually taken from  experiment as $\sqrt{K}=0.44 $GeV
where the value comes from $c \bar{c}$ and $b \bar{b}$ spectroscopy and
from the light  meson spectrum interpreted as excitations of a
relativistic string. Similar analyses also imply that $r_0 \approx 0.5 $
fm. Here we use $r_0^{-1}=0.372$ GeV  to be specific.  Since we shall be
describing quenched lattice results, the energy scale set from different
physical quantities will not  necessarily agree (since {\it experiment}
has full QCD not the quenched vacuum) and so a systematic error must be
applied to any such choice of scale. This, for instance, must be kept in
mind when taking glueball mass  values from the lattice.

The lattice potential $V(R)$ can be used to determine the spectrum
of $b \bar{b}$ mesons by solving Schr\"odinger's equation since the
motion is reasonably approximated as non-relativistic. The lattice result
is similar to the experimental $\Upsilon$ spectrum. The main
difference is that the Coulombic part ($e$) is effectively too
small (0.25 rather than 0.4). This produces~\cite{cmper} a ratio of
mass differences $(1P-1S)/(2S-1S)$ of 0.71 to be compared
with the experimental ratio of 0.78.  This difference is understandable
as a consequence of the Coulombic force at short distances which would be
increased by $33/(33-2N_f)$ in perturbation theory in full QCD compared
to quenched QCD. We will return to discuss this.

Another feature of the lattice determination of potentials is that  the
energy of static quarks at separation $R$ with an excited gluonic  field
can be determined.  This enables predictions to be made for  hybrid
mesons (eg. $b G \bar{b}$ where $G$ stands for the gluonic excitation).
Such mesons can have $J^{PC}$ values not allowed to $b \bar{b}$ states
and this allows them to be explored experimentally.  The current
situation~\cite{aachen,cmper} is  that such states, as predicted by the
lattice, will be  at high masses and hard  to isolate experimentally.

At small $R$, the static potential can be used, in principle,  to study
the running coupling constant. Small $R$ corresponds  to large momentum
and thus the coupling should decrease at  small $R$. Thus the Coulombic
coefficient $e$ introduced above should actually  decrease
logarithmically as $R$ decreases. Perturbation theory can  be used to
determine this behaviour of the potential at small $R$.

In the continuum the potential between static quarks is known
perturbatively to two loops in terms of  the  scale $\Lambda_{\msbar} $.
For  $SU(3)$ colour, the continuum force is given by \cite{bill}

\begin{equation}
{dV \over dR } =  {4 \over 3} {\alpha(R) \over R^2}
\end{equation}

\noindent with the effective coupling $\alpha (R)$ given by

\begin{equation}
 { 1 \over 4 \pi [ b_0  \log (R\Lambda _R )^{-2} +
(b_1 / b_0 ) \log \log (R\Lambda_R )^{-2} ] }
\end{equation}

 \noindent where $b_0=11/16 \pi ^2$ and $b_1=102 \ b_0^2/121$ are the
usual coefficients in the  perturbative expression for the
$\beta$-function, neglecting quark loops in the vacuum.
 Here  $\Lambda_R= 1.048 \Lambda_{\msbar}$.

On a lattice the force can be estimated by  a finite difference and one
can  extract  the  running  coupling constant by using~\cite{cmlms}

\begin{equation}
 \alpha( { \hat{R}_1 + \hat{R}_2 \over 2 }) = { 3\over 4} \hat{R}_1
\hat{R}_2 { \hat{V}(\hat{R}_1)-\hat{V}(\hat{R}_2) \over
\hat{R}_1-\hat{R}_2 }
 \end{equation}

\noindent where the error in using a finite difference is  negligible in
practice.

This  is plotted  in  fig~4 versus $R \sqrt K$: this combination  is
dimensionless and  so can be  determined from lattice results since $R
\sqrt K=\hat{R} \sqrt{ \hat{K}}$,  where $\hat{K}$ is taken from the fit
to $\hat{V}(\hat{R})$. The interpretation of $\alpha$ as defined above
as an effective running coupling constant is only justified at small $R$
where the  perturbative expression dominates. Also shown  are  the
two-loop perturbative results for $\alpha(R)$ for  different values of
$\Lambda_R $.

\begin{figure}[p]
\vspace{12cm} 
\includegraphics{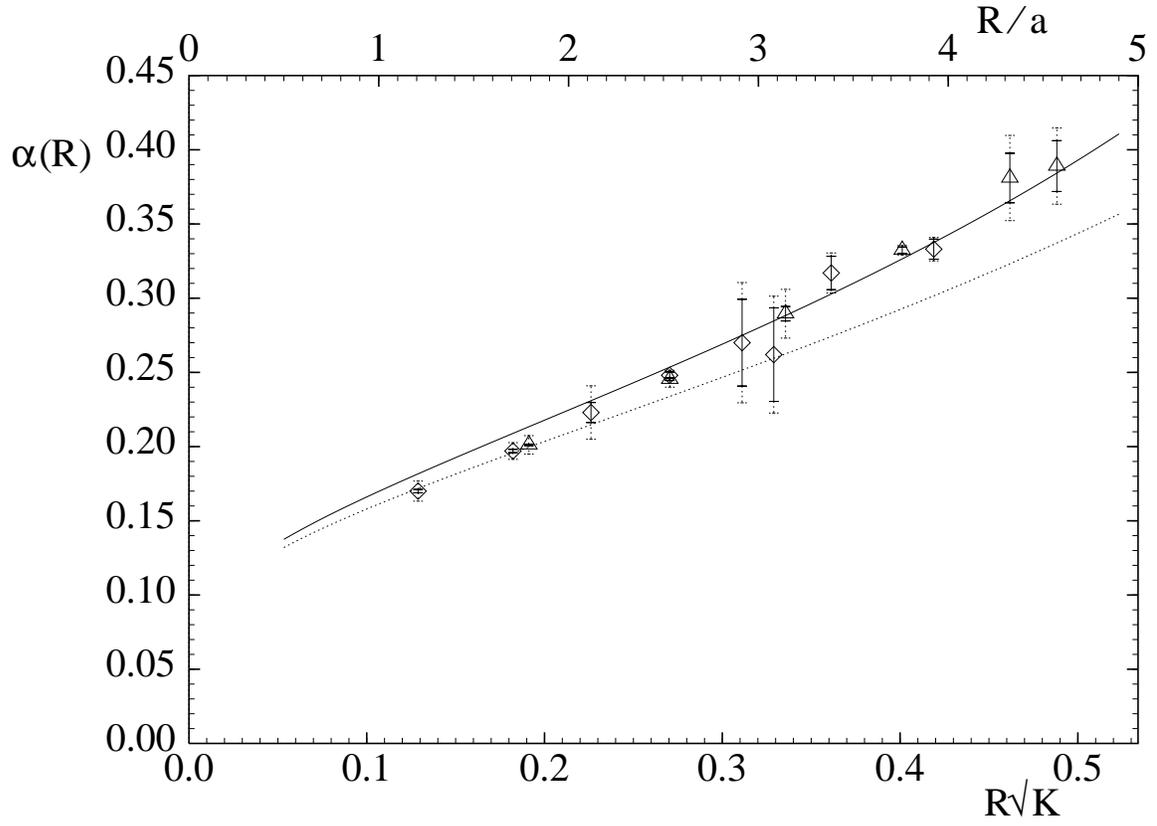}
\caption{
 The effective running coupling constant $\alpha(R)$ obtained from  the
force between static quarks at separation $R$ from
ref{\protect\cite{uksu3}}.  The scale is set by the string tension $K$.
Data are at $\beta=6.5$ (diamonds) and at $\beta=6.2$ (triangles).  The
dotted error bars represent an estimate of the systematic error due to
lattice artefact correction.  The curves are the two-loop perturbative
expression with $a(6.5)\Lambda_R=0.060$ (dotted) and 0.070 (continuous).
 }
\end{figure}

The figure clearly shows a {\it running} coupling constant.  Moreover
the result is consistent with the expected perturbative dependence on
$R$ at small $R$.  There are systematic errors, however. At larger $R$,
the perturbative two-loop expression will not be an accurate estimate of
the measured potentials, while at smaller $R$, the lattice artefact
corrections (which arise because $R/a \approx 1$) are
relatively big.  Setting the scale using $\sqrt K=0.44$ GeV implies
$1/a(\beta=6.5)=4.13 $ GeV, so $R < 4a(6.5)$ corresponds to values of
$1/R > 1$ GeV.  This $R$-region is expected to be adequately described
by perturbation theory.

This determination from the interquark force of the coupling $\alpha$
allows us to compare the result with the bare lattice coupling
determined  from $\beta=6/g^2$. At $\beta=6.5$, $\alpha_{\rm
bare}=g^2/4\pi=0.073$.  the values of $\alpha$ shown in fig~4 are much
larger. The effective  coupling constant is thus almost twice the bare
coupling. This is  quite acceptable in a renormalisable field theory.
The message is that  the bare coupling should be disregarded - it is not
a good expansion  parameter. The measured $\alpha$, however, proves to
be a reasonable  expansion parameter in the sense that  the first few
terms of the perturbation series converge.  This successful calibration
of  perturbation theory on a lattice is important in practice. For
instance,  when matrix elements are measured on a lattice they have
finite correction factors (usually called Z) to relate them to continuum
matrix elements. These  Z factors are evaluated perturbatively - so an
accurate continuum  prediction needs trustworthy perturbative
calculations.

\section{Full QCD}

So far we have  discussed  the glueball spectrum, interquark potentials
and $\alpha_S$ in the quenched approximation. This corresponds
to treating the sea quarks as of infinite mass (so they don't
contribute to the vacuum).  To make direct comparison with experiment,
it is necessary to estimate the corrections from these dynamical quark
loops in the vacuum.

The strategy is to use a finite sea-quark mass  but still a value larger
than the empirical light quark mass. The  reason is computational: the
algorithms become very inefficient as the  sea-quark mass is reduced.
The target is to study the effects as  the sea-quark mass is reduced and
then extrapolate to the  physical value.  The present situation,  in
broad terms, is that there is no significant change as the sea-quark
mass  is reduced. This could be because there are no corrections to  the
quenched approximation. Alternatively, the corrections may only turn  on
at a much lower quark mass than has been explored so far.

Let us try to make this argument a little more quantitative. For heavy
sea-quarks of mass $m$, their contribution will be approximately
proportional to $e^{-2m/E}$ where $E$ is a typical hadronic energy scale
(a few hundred MeV). Thus the quark loop contributions will be
negligible for $m >> E$, which corresponds to the  quenched
approximation. As $m \approx E$, the effects will turn on in a
non-linear way.

The computational overhead of full QCD on a lattice is  so large because
the quark loops effectively introduce a long range  interaction. The
quark interaction in the Lagrangian is quadratic  and so can be
integrated out analytically. This leaves an effective  Lagrangian  for
the gluonic fields which couples together the fields at all sites.
This implies that, in a Monte Carlo method, a change in gluon field
at one site involves the evaluation of the interaction with all other sites.
In practice, one makes small changes at all sites in parallel, but
this still amounts to inverting a large sparse matrix for each
update. This is computationally slow.

As the sea-quark mass becomes small, one would expect to need a larger
lattice  size to hold the quarks. For heavy quarks,  the effective range of
the quark loops  in the vacuum will be of order $1/m$. Thus the quenched
approximation  corresponds to $m \to \infty$ and a local interaction.
For light quarks of a few MeV mass, the range will not be $1/m$,
because quarks are confined. The lattice studies that have been made
suggest that spatial sizes of order twice those adequate for the quenched
approximation are needed for full QCD. This also implies considerable
computational commitment.

A popular indication of how close a full QCD study is to experiment
is to ask whether the $\rho$ meson can decay to two pions.  Since the
decay is P-wave, it needs non-zero momentum. On a lattice
spatial momentum is quantised in units of $2\pi/L$. Thus we need
$m_V > 2 m_P + 2\pi/L$ for the decay channel to be allowed energetically.
At present this criterion is rarely satisfied in quenched studies, let
alone in the more computationally demanding case of full QCD.

The  conclusion of current full QCD lattice calculations is that
the expected sea-quark effects are not yet fully present.
The main effect observed in full QCD calculations is
that the lattice parameter $\beta$ which multiples the gluonic
interaction term in the Lagrangian is shifted. Apart from this
renormalisation of $\beta$, there is little sign of any other
statistically significant non-perturbative effect.

Consider the changes to be expected for the  inter-quark potential when
the full QCD vacuum is used:
 \begin{itemize}
 \item At small separation $R$, the quark loops will increase the
size of the effective coupling $\alpha$ compared to the pure gluonic case.
This effect can be estimated in perturbation theory and the change
at lowest order will be from 1/33 to 1/(33-$2N_f$).
 \item At large separation $R$, the potential energy will saturate at
a value corresponding to two `heavy-quark mesons'. In other
words, the flux tube between the static quarks will break by the
creation of a $q \bar{q}$ pair from the vacuum.
 \end{itemize}
 Current lattice simulation~\cite{dynV} shows some evidence for the
former effect  but no statistically significant signal for the latter.

If one  assumes that these lattice simulations are an approximation  to
the true full QCD vacuum, then one can use them to estimate  the  full
QCD running coupling from lattice studies. A summary~\cite{cmlat94} of
present  lattice results is that
 $$
\alpha_{\msbar}(M_Z) =0.112(7)
 $$
This conclusion will be reinforced when the full QCD lattice
results reproduce more of the features of the experimental
spectrum.

\section{Quenched hadron masses}

Since the full QCD simulation is not feasible at present, it is
worthwhile to explore the hadron spectrum in the quenched
approximation. This amounts to allowing quarks to propagate  in the
gluonic vacuum. Computationally this can be studied  by solving the
lattice Dirac equation for the quarks. Since the  gluonic vacuum is full
of rich structure, this is a computationally  intensive problem: it
amounts to inverting a large sparse matrix. Indeed it proves necessary
to compute the quark propagation for  a range of valence quark masses
larger than the physical light  quark masses and then extrapolate.
Because of this, the statistical precision of such calculations   is
still somewhat limited. Nevertheless, different groups  using different
methods agree on the main results. For a recent review see
ref\cite{cmlat94}.

Let us first discuss the $\rho$ meson. In fig~5, the mass is  compared
{}~\cite{soml94} to a lattice scale (here $\Sigma=Ka^2$ where $K$ is the
string tension). The figure shows that different treatments of fermions
on the lattice  (Wilson, Clover and staggered)  with different
discretisation errors ($a$, $\alpha_S a$ and $ a^2$ respectively) are
in agreement with a common continuum  limit. Moreover, using the usual
convention that the string tension is $\sqrt{K}=0.44$ GeV, the continuum
value of $m_{\rho}/\sqrt{K}$ of 1.80(5) yields $m_{\rho}=0.79(2)$ GeV
which is consistent with the experimental value of 0.77 GeV.

Studies have been made of the mesons and baryons which are composed of
light and strange quarks- see ref\cite{cmlat94}. The surprise is that
the quenched approximation  seems to reproduce these experimental values
quite well. This may  reflect the relatively large errors that are still
present in the  lattice determinations. It may also reflect the fact
that the hadronic  dynamics has a similar energy scale in each case so
that the quenched  approximation makes similar errors - which cancel in
mass ratios.

\begin{figure}[p]
\vspace{11.6cm}  
\includegraphics{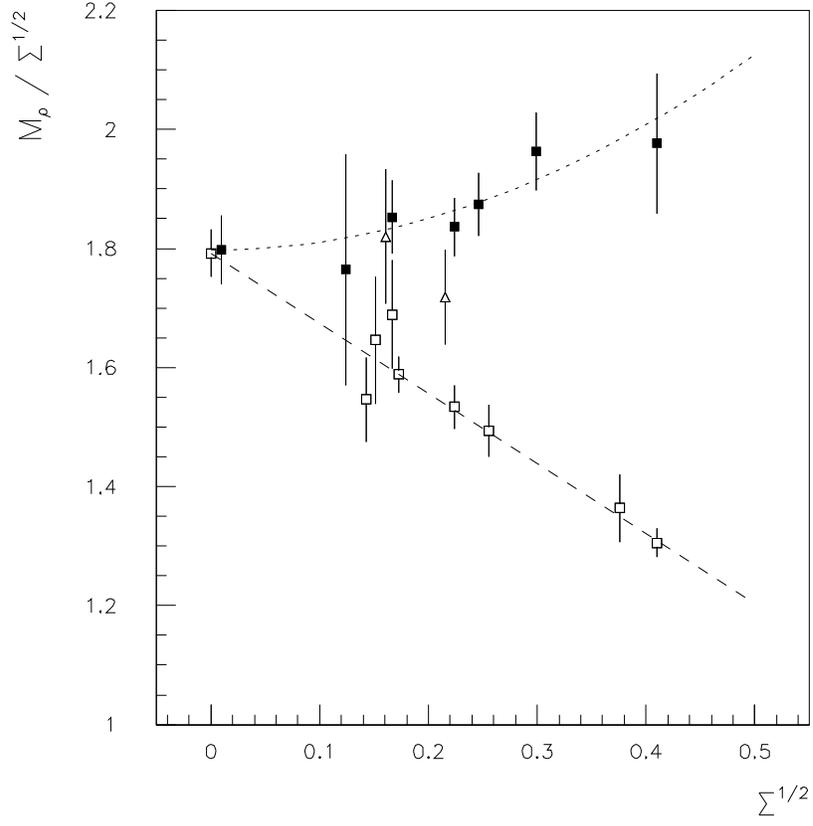} 
\caption{
 The mass of the $\rho$ meson as a dimensionless ratio to  the lattice
string tension $\Sigma=Ka^2$. The continuum limit corresponds  to $a \to
0$ at the left hand side. The filled squares are with  staggered
fermions, the open squares from Wilson fermions and  the triangles are
from Clover fermions. The lines show the discretisation  errors which
behave as $a^2$, $a$ and $\alpha_s a$ respectively.
 }
\end{figure}

\section{Matrix elements eg $f_P$}

\begin{figure}[t]
\vspace{8cm}
\includegraphics{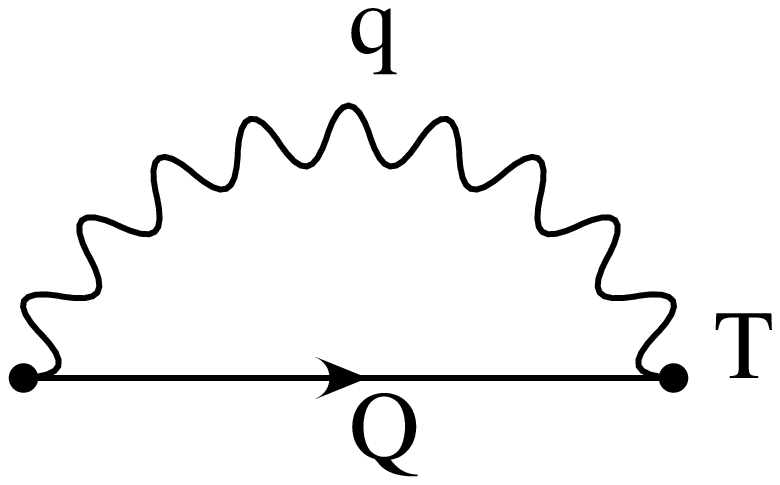}
  \caption{ The diagram representing a static quark Q propagating in
the time direction and a light quark q which creates a gauge invariant
quantity appropriate for studying $f_B$.
   }
\end{figure}

One of the advantages  of lattice QCD is that it is a method to
calculate hadronic matrix elements from first principles. Consider as an
example the  weak decay of a pseudoscalar meson P. The weak axial
current will  couple to the quarks in the meson. This current will be
local. Thus  the required quantity will relate the quark current to the
hadronic  state P.  This is the pseudoscalar decay constant $f_P$
which is defined by the matrix element of the divergence of
the axial current. For a pseudoscalar state of zero spatial momentum,
 $$
<0| q_{\mu} j^A_{\mu} |P> = {f_P \over \sqrt{2 m_P} }
 $$
 where $j^A$ is the axial current which in terms of quark  fields is
$\bar{q} \gamma_5 \gamma_{\mu} q$. For the pion, this identity is the
partially  conserved axial current (PCAC) relation and $f_{\pi}$ is the
coupling of the  pion to the weak current (and hence is relevant to the
$\mu+\nu$ decay mode of the pion). Since the axial current is
represented by local quark fields, $f_P$ gives  a relationship between
the hadronic state ($|P>$) and the quark  sub-structure ($j^A$).

For a pseudoscalar meson with a heavy quark, such as the B meson, this
same relationship is needed. Because the relevant weak decay is not
currently observable (branching ratio too small), a lattice calculation
is needed to determine $f_B$. The picture, for a heavy quark, is
clarified  by the heavy quark effective theory (HQET) which treats the
heavy  quark as slow moving. In the static limit, the lattice
calculation needed  is easily visualised. A straight line of colour flux
of length $T$ in the  lattice time direction represents the propagation
of the heavy quark. By  combining this with the propagator from one end
to the other of a light quark in the lattice gluonic  background field,
one has  a gauge invariant quantity which can be measured on a lattice
(see fig~6). At each end, one joins the heavy quark to the light  quark
by a local axial current ($\gamma_5 \gamma_4$ for a state at rest). Then
the observed correlation $C(t)$ is proportional to $f_P^2 e^{-E_P T}$
for  large $T$ so allowing $f_P$ and $E_P$ do be determined in
principle.  In practice more sophisticated methods are used to improve
the lattice  measurement signal.  Here $E_P$  has the interpretation of
the mass difference of the B meson and the  $b$ quark - although this
difference is not directly useful since  a non-perturbative definition
of the quark mass requires careful discussion.

As well as using static quarks, lattice studies have been performed
using propagating heavy quarks. Combining the results from both
studies~\cite{soml94}
yields $f_B=180(50)$ MeV. This value is needed as an ingredient  in using
experimental data to fix  the CKM weak matrix elements. This is turn has
implications for the  experimental feasibility  of CP violation  through
studies of $\bar{B} B $ mixing.

\section{Outlook}

Lattice techniques can extract reliable continuum properties from QCD.
At present, the computational power available combined with the best
algorithms suffices to give accurate results for many quantities  in the
quenched approximation. The future is to establish accurate  values for
more subtle quantities in the quenched approximation (eg. weak matrix
elements of  strange  particles) and to establish the validity  of the
quenched approximation by full QCD calculations.

We need to reach the stage where an experimentalist saying `as
calculated in QCD' is assumed to be speaking of non-perturbative
lattice calculations rather than perturbative estimates only.

\end{document}